\documentclass[aps,pra,superscriptaddress,12pt,tightenlines,nofootinbib]{revtex4}
\usepackage{amsmath,amsthm,graphicx,amssymb,verbatim,listings}
\usepackage{wasysym}
\usepackage[T1]{fontenc}     % needed for the guillemets
\usepackage{lmodern}         % needed to fix suboptimal font rendering
\usepackage[ pdftex, plainpages = false, pdfpagelabels,
                 pdfpagelayout = useoutlines,
                 bookmarks,
                 bookmarksopen = true,
                 bookmarksnumbered = true,
                 breaklinks = true,
                 linktocpage=all,
                 pagebackref=false,
                 colorlinks = true,
                 linkcolor = BrickRed,
                 urlcolor  = blue,
                 citecolor = BrickRed,
                 anchorcolor = green,
                 hyperindex = true,
                 hyperfigures
                 ]{hyperref}
\usepackage[usenames, dvipsnames]{xcolor}
\usepackage{xifthen} % provides \isempty test

\newcommand{\tr}{\hbox{tr}}

\newcommand{\ket}[1]{{\ensuremath{\left| #1 \right\rangle}}}
\newcommand{\bra}[1]{{\ensuremath{\left\langle #1 \right|}}}
\newcommand{\braket}[2]{{\ensuremath{\left\langle #1 \middle| #2
      \right\rangle}}}
\newcommand{\ketbra}[2]{{\ensuremath{\left| #1 \middle\rangle\!\middle\langle #2
      \right|}}}

\newcommand{\inprod}[2]{{\left\langle #1 , #2 \right\rangle}}
\newcommand{\varinprod}[2]{{\left( #1 , #2 \right)}}

\newcommand{\qbar}{\ensuremath{q\hspace{-1.10ex}{\rule[-0.255ex]{0.45em}{0.09ex}}}}

\newcommand{\arxiv}[2][]{\ifthenelse{\isempty{#1}}{\href{http://arxiv.org/abs/#2}{{\tt arXiv:\allowbreak{}#2}}} {\href{http://arxiv.org/abs/#2}{{\tt arXiv:\allowbreak{}#2 [#1]}}}}
\newcommand{\pirsa}[1]{\href{http://pirsa.org/#1/}{{\tt PIRSA:\allowbreak{}#1}}}
\newcommand{\booktitle}{\textsl}
\newcommand{\hrefdoi}[2]{\href{https://dx.doi.org/#1}{#2}}

\newcommand{\cA}{\mathcal{A}}
\newcommand{\cP}{\mathcal{P}}
\newcommand{\cR}{\mathcal{R}}

\newcommand{\cH}{\mathcal{H}}

\newcommand{\bbR}{\mathbb{R}}
\newcommand{\bbC}{\mathbb{C}}
\newcommand{\bbH}{\mathbb{H}}

\newcommand{\cone}{\mathcal{C}}
\newcommand{\orderunit}{\mathcal{I}}

\begin{document}

\title{Quantum Theory as Symmetry Broken by Vitality}

\author{Blake C.\ Stacey}
\affiliation{\href{http://www.physics.umb.edu/Research/QBism/}{QBism Research Group}, University of Massachusetts Boston}

\date{\today}

\begin{abstract}
I summarize a research program that aims to reconstruct quantum theory
from a fundamental physical principle that, while a quantum system has
no intrinsic hidden variables, it can be understood using a
\emph{reference measurement}. This program reduces the physical
question of why the quantum formalism is empirically successful to the
mathematical question of why complete sets of equiangular lines appear
to exist in complex vector spaces when they do not exist in real
ones. My primary goal is to clarify motivations, rather than to
present a closed book of numbered theorems, and consequently the
discussion is more in the manner of a colloquium than a PRL.
\end{abstract}
\maketitle

\section{Introduction}
I have been looking up science shows I saw as a kid and rewatching
them, partly to see how well their content has aged, and partly as
``relaxation tapes'' to help me decompress. On occasion, a bit will
jump out at me and become particularly relevant to my current
interests. The two examples I have in mind right now arose from the
Public Broadcasting System of ages past, so what follows was made
possible by the generosity of Viewers Like You. First, there's
David L.\ Goodstein in \booktitle{The Mechanical Universe
  and Beyond} (1986):
\begin{quotation}
\noindent The science of thermodynamics is based on four fundamental
postulates, or axioms, which are called the Four Laws of
Thermodynamics. Of these four laws, the second law was discovered
first, and the first law was discovered second, and the third to be
discovered was called the zeroth law, and the fourth law is called the
third law. Now, all of that makes perfect sense because thermodynamics
is the most implacably logical of all the
sciences~\cite{Goodstein:1986}.
\end{quotation}
The second example is Timothy Ferris in \booktitle{The Creation of the
  Universe} (1985):
\begin{quotation}
\noindent Perfect symmetry may be beautiful, but it's also sterile.
\end{quotation}

In this essay, my conceit will be that there are Laws of Quantum
Mechanics, analogous to the Laws of Thermodynamics and to Einstein's
postulates for special relativity. I will try to develop the thesis
that \emph{quantum physics is the simplest possible theory of
probability for a world where there are no intrinsic hidden
variables.} Of course, giving meaning to a claim like that requires
getting precise about terms like \emph{simplest.} This claim can
indeed be made precise enough to be mathematical. While we cannot
prove it, we can reduce it reasonably well to a few well-posed
conjectures.

I will attempt to explain why particular statements ought to be true,
not necessarily proving them yet in all detail. My goal is to lay out
my motivations and heuristics, rather than to make a closed book of
definitions and lemmas and corollaries. My approach will be grounded
in previous QBist and QBist-adjacent writing on the reconstruction of
quantum theory. Chiefly, this means our paper ``Introducing the
Qplex''~\cite{Appleby:2017}, and a line of thinking that had been
confined to an appendix therein.
\begin{enumerate}
\setcounter{enumi}{-1}
\item Two states of expectation are equivalent for an arbitrary
  measurement when they are equivalent for the reference measurement.
\item Certainty is achievable, and it defines the boundary of state
  space. In particular, the states most closely associated with the
  reference measurement each imply certainty for some experiment.
\item Yet certainty is \emph{not} about hidden
  variables:\ \emph{Unperformed experiments have no results.}
\item As in classical probability, states in the interior of the state
  space can be reversibly mapped to one another.
\end{enumerate}

Asher Peres' slogan, ``Unperformed experiments have no
results''~\cite{Peres:1978}, is typically said to summarize the fact
that ``no-go theorems'' rule out hidden-variable completions of
quantum mechanics. It is convenient to have a positive expression of
what is normally stated negatively, and so we will entertain a
counterpart slogan, \emph{quantum physics has
  vitality}~\cite{Stacey:2016}.

The \emph{physical} assumption underlying all of this development is
the possibility of expressing this vitality using a \emph{reference
  measurement}. We have taken this route before, but compared to what
I would like to do, our prior work jumped somewhat into the middle,
and it did not fully embrace the theme that I intend to explore
now. If there is truly one fundamental mystery and all the rest of the
quantum formalism is mathematical niceties, then perhaps those should
be expressed as mathematicians do --- that is, by talk of symmetries,
and the properties left invariant by transformations. Perfect symmetry
is sterile; all the life lies in the fundamental axiom. Therefore, all
the later steps in the derivation of the quantum follow the route of
maximal symmetry.

A conceptual tool that I will use to narrow the mathematical
possibilities is van Fraassen's \emph{reflection principle,} which
gives meaning to convex combinations of probabilities and provides a
reason to believe in linearity~\cite{Goldstein:1983, Shafer:1983,
  VanFraassen:1984, Fuchs:2012}.

\section{Preliminaries and Definitions}
\begin{flushright}
When I read someone else's math,\\
I always hope the author\\
will have included a reason \\
and not just a proof.\\
\smallskip
--- Eugenia Cheng~\cite{Cheng:2015}
\end{flushright} 

Our goal is the finite-dimensional quantum theory familiar from
quantum information and computation. That is, we wish to explain why
applying the theory means associating a physical system with a complex
Hilbert space, why positive semidefinite operators on that Hilbert
space are so important for multiple purposes, why the textbook rule
for calculating probabilities is the correct one, and so on. Efforts
to put quantum theory on a more principled foundation are almost as
old as quantum physics itself. However, many of these research
endeavors antedate the mature understanding of what the most enigmatic
features of quantum physics truly are. Even the modern renaissance of
quantum reconstructions, beginning around the turn of this century,
has largely been preoccupied with making the theory seem as ``benignly
humdrum'' as possible~\cite{Fuchs:2016}. Rather than starting with a
remarkable phenomenon --- say, the violation of a Bell inequality ---
and building the subject up from there, the reconstructors' ethos has
mostly been to draw up lists of postulates that, individually, sound
thoroughly pedestrian. From such a list, the quantum formalism is
rederived. And then, given that formalism, the remarkable features can
be exhibited as they had been before. The undeniable strangeness of
quantum phenomena is not written in any one axiom, but somehow
interleaved between them, sometimes in the tacit conditions accepted
without demur but not given bold text and bullet points. The bolded
axioms themselves are often satisfied by fundamentally classical
theories, like the Spekkens toy model~\cite{Spekkens:2007}! Thus,
while the mathematics may be sound --- the density of errata is
probably no worse than average for the \booktitle{Physical Review}
family --- the result is vaguely dispiriting all the
same. Reconstructing the quantum was supposed to free us from
antiquated debates over ``interpretations'', but how can it do that if
the lists of new operational postulates themselves multiply like the
``interpretations'' have?

Accordingly, while we will lean upon the algebraic achievements of
these reconstruction efforts, we will arrive at the point where that
algebra can be invoked in our own way.

The present work is firmly in the personalist Bayesian
tradition. Apart from some changes in emphasis, the QBist view on how
to interpret probability is closely kin to that espoused by Diaconis
and Skyrms~\cite{Diaconis:2018}, and it has been developed over
several previous publications~\cite{Fuchs:2012, Fuchs:2009,
  Fuchs:2013b, Fuchs:2013c}. Elsewhere, I have built up the theory
pedagogically to the point where one can do nonequilibrium statistical
physics with it~\cite{Stacey:2015, Stacey:2025}. Well before I acquired a serious
interest in quantum foundations, I had a nudge in this general
direction, thanks to taking statistical mechanics from Mehran
Kardar. To quote his \booktitle{Statistical Physics of
  Particles}~\cite{Kardar:2007}, ``All assignments of probability in
statistical mechanics are subjectively based.'' (For all its talk of
``ensembles'', statistical mechanics really gives lessons in how to
adopt particular priors~\cite{Rau:2017}. Indeed, as the concepts are
applied in practice, the term ``ensemble'' becomes less and less an
appeal to relative frequencies, and more a meat-noise emitted out of
habit.) If anyone is aghast at the spectacle of a youth turning to
QBism, well, they can blame the Massachusetts Institute of Technology.

Topology is the study of properties invariant under continuous
transformations, and differential geometry studies quantities that are
invariant under coordinate changes. Terry Tao suggests that we should
think of probability theory analogously, as studying those concepts
and operations that are preserved by extending sample
spaces~\cite{Tao:2012}. This is not dissimilar to the Dutch-book view,
where one imposes the rule that two propositions should be ascribed
the same probability if they are equivalent by Boolean
grammar. Moreover, this motivates the idea that we should be able to
apply our theory to systems of arbitrary dimensionality, once we have
properly made precise what ``dimensionality'' means.

Let $d$ be an integer greater than 1, and consider the vector space
$\bbC^d$. A \emph{SIC} is a set of $d^2$ unit vectors
$\{\ket{\pi_j}\}$ in this space which enjoy the property that
\begin{equation}
|\braket{\pi_j}{\pi_k}|^2 = \frac{d\delta_{jk} + 1}{d+1}.
\end{equation}
The acronym SIC (pronounced ``seek'') stands for ``Symmetric
Informationally Complete'' and refers to the fact that such a set
represents a measurement that can be performed upon a quantum system
of Hilbert-space dimension $d$~\cite{Zauner:1999, Renes:2004,
  Scott:2010, Scott:2017}.  Because the operators $\Pi_j =
\ketbra{\pi_j}{\pi_j}$ span the space of Hermitian operators on
$\bbC^d$, any quantum state $\rho$ that one might ascribe to the
system can be expressed in terms of its inner products with those
operators, which up to normalization are just the probabilities for
the outcomes of the SIC measurement:
\begin{equation}
p(j) = \frac{1}{d} \tr(\rho\Pi_j).
\end{equation}
Prior work has established the virtues of SICs as reference
measurements~\cite{Fuchs:2013b, Zhu:2016, DeBrota:2018a,
  DeBrota:2018b}. They are also of interest for purely mathematical
reasons~\cite{Bengtsson:2017, Appleby:2017b, Stacey:2017,
  Appleby:2017c, Fuchs:2017, Kopp:2018}. SICs are the largest possible
sets of equiangular lines; that is, one cannot have more than $d^2$
unit vectors in $\bbC^d$ such that $|\braket{\pi_j}{\pi_k}|$ is
constant for all $j \neq k$. As of this writing, SICs are known
numerically for all $d$ up to 193, and in irregular cases up to $d =
39604$, while exact solutions have been found for all $d$ up to 53 and
irregularly up to $d = 5799$~\cite{Appleby:2019, Appleby:2023}.

Now, we have defined enough terminology to know what we mean when we
say, for example, ``Bengtsson, Blanchfield and
Cabello~\cite{Bengtsson:2012} proved a quantum vitality theorem using
a SIC in dimension 3.'' For the remainder of this section, we will
establish a little more jargon that will turn out useful.

A \emph{qplex} is a subset of the probability simplex of normalized,
entrywise nonnegative vectors in $\bbR^{d^2}$. Any two points within a
qplex are said to be \emph{consistent} in that their Euclidean inner
product satisfies the inequalities
\begin{equation}
\frac{1}{d(d+1)} \leq \inprod{p_1}{p_2} \leq \frac{2}{d(d+1)}.
\end{equation}
Moreover, a qplex is a \emph{maximal} consistent set:\ If one tries to
introduce even a single additional point, inconsistencies arise.  A
\emph{Hilbert qplex} is one whose symmetry group is isomorphic to the
projective extended unitary group $\mathrm{PEU}(d)$. Hilbert qplexes
are images of quantum state spaces under the mappings defined by SICs;
the above inequalities are the image of the statement that $\tr
\rho_1\rho_2$ lies in the unit interval for any two quantum states
$\rho_1$ and $\rho_2$. A primary goal of this essay is to replace the
assumption of projective unitary group symmetry with a postulate that
is less specific but just as powerful.

It follows from the maximality property that a qplex is necessarily
convex and closed. Thanks to Huangjun Zhu, we know many more things
about qplexes. For instance, we know that any qplex is a
\emph{self-polar} set. Let $H$ be the hyperplane in $\bbR^{d^2}$
consisting of vectors whose elements sum to unity, i.e., the
hyperplane of probabilities and quasiprobabilities. The polar of a
point in~$H$ is the set of all points in~$H$ whose inner product with
the given point is greater than the lower bound in the fundamental
inequalities. The polar of a set of points is the set of all points
which are in the polars of all the given points. (This terminology is
adapted from the study of polytopes.) The operation of taking the
polar reverses inclusion, so the polar of a set that lies within the
probability simplex contains the polar of the probability simplex,
which is another simplex whose vertices are the probability
distributions
\begin{equation}
  e_j(i) = \frac{1}{d+1}\delta_{ij} + \frac{1}{d(d+1)}.
\end{equation}
Note that the constant term is just the lower bound. We refer to these
vectors as the \emph{basis distributions}. When considered together,
they form a matrix whose inverse is
\begin{equation}
  \Phi = (d+1)I - \frac{1}{d}J, \label{eq:Phi}
\end{equation}
using $J$ to stand for the all-ones matrix (the Hadamard identity).

Another important fact about qplexes is the maximal size of a
\emph{mutually maximally distant} (MMD) set. Let
$\{p_j:j=1,\ldots,m\}$ be a set of points in a qplex, such that
$\inprod{p_j}{p_j}$ equals the upper bound and, when $j \neq k$,
$\inprod{p_j}{p_k}$ equals the lower bound. Such a set can only be so
big. In fact, $m \leq d$.
  
A \emph{generalized qplex} is defined similarly as a maximal
consistent set of probability vectors in $\mathbb{R}^N$, where
consistency is with respect to the inequalities
\begin{equation}
  L \leq \inprod{p_1}{p_2} \leq U.
\end{equation}
Polarity works much as before in this more general setting. The basis
distributions are the vectors
\begin{equation}
e_k(j) = (1-NL)\delta_{jk} + L,
\end{equation}
again found by taking all entries save one to be the lower bound. The
matrix $\Phi$ that we defined in Eq.~(\ref{eq:Phi}) generalizes to
\begin{equation}
  \Phi= \frac{1}{1 - NL}(I - LJ). \label{eq:gen-Phi}
\end{equation}
It will be helpful later to note that
\begin{equation}
  \inprod{p_1}{\Phi p_2} = \frac{1}{1-NL}\inprod{p_1}{p_2 - LJp_2}
  = \frac{1}{1-NL}(\inprod{p_1}{p_2} - L).
\end{equation}
One intriguing property enjoyed by generalized qplexes is that the
number of zeros in any probability vector cannot exceed $N - 1/U$.

\section{Managing Expectations in a Reality Too Rich for Turing Machines}

In earlier work, I have drawn a connection between SICs and
thermodynamically significant quantities~\cite{Stacey:2019b}. Here, I
want instead to make a much higher-level argument, following the
analogy that Chris Fuchs and I articulated in our ``Hero's Handbook''
paper~\cite{Fuchs:2019}:
\begin{quotation}
\noindent We can illustrate the trouble with quantum mechanics by
comparing it with other areas of physics in which we have collectively
honed our understanding to a high degree of sophistication.  Two
examples that come to mind are the science of thermodynamics and the
special theory of relativity.  An old joke has it that the three laws
of thermodynamics are ``You can't win'', ``You can't break even'', and
``You can't get out of the game.''  To these, we ought to prepend the
zeroth law, which we could state as, ``At least the scoring is fair.''
But consider the premise of the joke, which is really rather
remarkable:\ There \emph{are\/} laws of thermodynamics --- a concise
list of deep physical principles that underlie and nourish the entire
subject.  Likewise for special relativity: Inertial observers Alice
and Bob can come to agree on the laws of physics, but no experiment
they could ever do can establish that one is ``really moving'' and the
other ``really standing still'' --- not even measuring the speed of
light.  We invest a little mathematics, and then close and careful
consideration of these basic principles yields all the details of the
formal apparatus, with its nasty square roots or intermingling partial
derivatives.

This level of understanding brings many advantages.  Having the deep
principles set out in explicit form points out how to \emph{test\/} a
theory in the most direct manner.  Moreover, it greatly aids us when
we \emph{teach\/} the theory.  We do not have to slog through all the
confusions that bedeviled the physicists who first developed the
subject, to say nothing of the extra confusions created by the fact
that ``historical'' pedagogy is almost inevitably a caricature.  In
addition, a principled understanding helps us \emph{apply\/} a theory.
As we work our way into a detailed calculation, we can cross-check
against the basic postulates.  Does our calculation imply that signals
travel faster than light?  Does our seventeenth equation imply that
entropy is flowing the wrong way?  We must have made an error!  And,
when we found our theory upon its deep principles, we have a guide for
\emph{extending\/} our theory, because we know what properties must
obtain in order for a new, more general theory to reduce to our old
one in a special case.

To our great distress, we must admit that in the matter of quantum
mechanics, the physics profession lacks this level of understanding.
\end{quotation}

We can make this analogy more exact~\cite{Stacey:2018}.  Both in
special relativity and in thermodynamics, one of the postulates has
the character of a guarantee: Inertial observers Alice and Bob can
come to agree on the laws of physics; energy is conserved. Then comes
a postulate that serves as a dramatic foil to the first, verging on
contradicting it.  No matter what they do, Alice and Bob cannot agree
on a standard of rest, even if they go so far as measuring the speed
of light. And in thermodynamics, energy is conserved, but
\emph{useful} energy diminishes in all but the most ideal
processes. In special relativity, we derive a statement of
unattainability: Massive bodies cannot attain light speed. Meanwhile,
over in thermodynamics, we assume a form of unattainability in the
Third Law.

What about the Zeroth Law? While the ordinary statement of it
disguises this a little, the way we use it in practice indicates that
it is a statement that holds moment-to-moment. At each instant of a
``quasi-static'' evolution, a system can be taken to be in equilibrium
with a fiducial heat bath at the appropriate temperature. This is
analogous to the clock postulate, the rule which Einstein invoked but
did not grant a number. Just as the moment-to-moment statement of the
Zeroth Law implies that we can use temperature as a reference scale
even during a nontrivial time evolution, the clock postulate lets us
use \emph{velocity} as a reference scale, analyzing accelerated motion
using a series of momentarily co-moving inertial frames.

Thus, we have a pedagogical schema that applies to both theories:
\begin{enumerate}
\setcounter{enumi}{-1}
\item Equivalence relation, implying a scale of reference
\item Reassuring guarantee
\item Dramatic foil with metaphysical weight
\item Unattainability, of an asymptotic flavor
\end{enumerate}

What, then, about quantum mechanics?

\begin{figure}[t]
    \resizebox{8.5cm}{4.5cm}{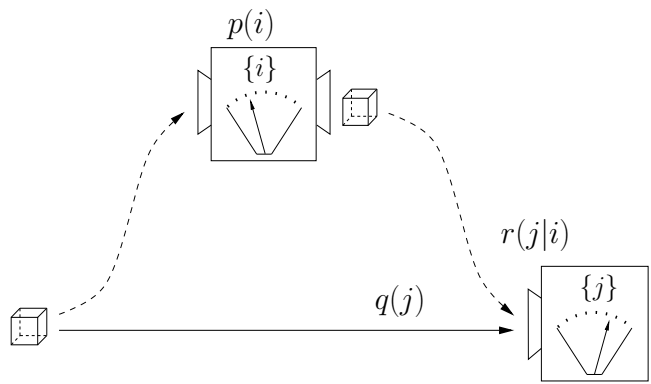}
    \caption{A choice between two experiments. In one scenario (solid
      line), a system is fed directly into a measuring appartus. In
      the other (dashed line), the system is sent through the
      reference measurement first. Probability theory does not itself
      enforce a relation between an agent's probabilities for these
      two scenarios. Different conditions, different probabilities!
      The classical intuition that the reference measurement just
      reads off the system's intrinsic degrees of freedom leads to
      using the Law of Total Probability to relate expectations
      between the two scenarios. Quantum theory, on the other hand,
      provides its own relation. Our goal is to identify exactly what
      physical principle implies the quantum
      relation. (After~\cite{DeBrota:2018a}.)}
\label{figure1}
\end{figure}

The view developed in the ``Hero's Handbook'' review~\cite{Fuchs:2019}
is that systems can have an arbitrarily rich supply of physical
properties, but these attributes do not, either singly or in
combination, compel the outcomes of measurements, or even the
probabilities that an agent should ascribe to them.  The system
attribute to which \emph{the basic quantum formalism} is sensitive is
the ``creative capacity'' that manifests as Hilbert-space dimension,
or the quantity in a more ambitious theory that reduces to it.

The ``zeroth law of quantum theory'' proposed above tells us that we
can think in terms of a reference measurement. In any situation, we
can calculate the probabilities we need in terms of the probabilities
$\{p(i)\}$ for the possible outcomes of the reference measurement, and
the conditional probabilities $\{r(j|i)\}$ for the result of our other
experiment given an outcome of the reference measurement:
\begin{equation}
  q(j) = \mu(p, r),
\end{equation}
where $p$ is a vector and $r$ is a matrix. The number of outcomes
necessary for a reference measurement, $N$, is regarded as an
intrinsic property of a physical phenomenon.  It is how the quantity
that we evocatively described as ``creative capacity'' enters the
theory. Our goal now is to characterize the function $\mu$. Note that
classically --- or perhaps better put, according to a sentiment of
classicality more primitively rooted than classical mechanics --- an
ideal ``reference measurement'' would simply read off the system's
intrinsic physical degrees of freedom, and we would have
\begin{equation}
q(j) = \sum_i p(i) r(j|i).
\end{equation}
This formula is the \emph{Law of Total Probability,} and it expresses
the intuition that classical uncertainty is ignorance of a system's
``physical condition'' (to use Einstein's
terminology~\cite{Stacey:2018b}) or ``ontic state'' (in a more modern
turn of phrase). We will show that the function $\mu$ cannot be of
this form, and that the correct quantum form of $\mu$ follows from an
expression of quantum vitality.

To vary our vocabulary, we will sometimes speak of the ``preparation''
of a system rather than the ``state'' ascribed to it. This is with the
understanding that, for example, picking up a rock from the beach is a
``preparation''.  We make no claim of expense or exactitude; a
preparation has the same status as a prior, with all the personalism
inherent in that. For us, \emph{preparation} is a synonym for
\emph{prior} with different scansion. Likewise, we call two
preparations ``distinguishable'' if they imply sharply discrepant
predictions for some experiment.

Other approaches have been criticized for saying that the laws of
physics ``state by fiat that a particular type of behavior is
rationally compulsory for rational agents''~\cite{Kent:2009}. We are
aiming for something more subtle and robust.  In our approach, we
acknowledge that we bring to the table certain choices --- about how
to represent the reasoning process, what mathematical entities can
stand for intensities of belief, etc. --- and we then manage our
expectations \emph{in light of} the natural world and its character.

The most arbitrary-looking mathematical decision occurs where the
rubber meets the road, i.e., where the character of natural pheomena
must find a representation within our conventions. For our purposes,
this means the way we choose to encode the nonexistence of intrinsic
hidden variables. In a classical theory, a reference measurement would
at the simplest just read off the intrinsic ``physical condition'' of
the system being measured. All other measurements could be understood
as coarse-grainings of this reference measurement, which we can
represent probabilistically as a filtering by a stochastic
process. Even if the reference measurement were not feasible in
practice, we could tie together our probabilities for those
experiments that \emph{are} practical, introducing a reference
experiment of which they are all post-processings, without fear of
inconsistency. One consequence of this is that if two preparations are
distinguishable with respect to some measurement, they must
also be so by the reference measurement. Post-processing by a
stochastic coarse-graining cannot eliminate an overlap between
probability distributions.

We posit that \emph{rejecting} this statement of how
distinguishability behaves captures a deep truth about quantum
physics. Two orthogonal quantum states are perfectly distinguishable
with respect to some experiment, yet in terms of the reference
measurement, they are \emph{inevitably overlapping} probability
distributions. The idea that any two valid probability distributions
for the reference measurement must overlap, and that the minimal
overlap in fact corresponds to distinguishability with respect to some
other test, expresses the fact that quantum probability is not about
hidden variables. Or, to make the statement in a positive form, it
expresses quantum vitality.  Thus the inequality
\begin{equation}
\inprod{p_1}{p_2} > 0
\end{equation}
looks small but packs a big punch.

There may be many possible reference measurements.  In fact, in
quantum theory, the set thereof is rather richly
structured~\cite{DeBrota:2018a, DeBrota:2018b}.  However, the
\emph{canonical} choice, the analogue of a Carnot engine or a light
clock, is a measurement that expresses vitality in the crispest
possible way. Therefore, we will take as canonical reference
measurement one for which there is a single constant value of overlap
that corresponds to perfect distinguishability:
\begin{equation}
\inprod{p_1}{p_2} \geq L.
\end{equation}

Imagine building up a qplex $\cP$ one point at a time. If we include a
point $p$, we automatically exclude all points that are too far
away from~$p$, that is, all vectors $v$ such that $\inprod{p}{v} <
L$.  Are there points that are \emph{never} excluded during this
process? Such probability distributions would be a necessary part of
\emph{any} qplex, for without them, the set could never be
maximal. The points that cannot be excluded are those $p$ for which
$\inprod{p}{v} \geq L$ for all $v$ in the probability simplex, or in
other words, the polar of the probability simplex.  Thus, we know that
all qplexes must include the convex hull of the basis distributions.
They tumble out once we introduce the idea of taking a polar, which
follows directly from our way of expressing that intrinsic hidden
variables don't exist!

Let $\mu_E: \cP \to [0,1]$ be the function that computes the
probability of event $E$ given a probability vector over the outcomes
of the reference measurement. Suppose that $p_1,p_2 \in \cP$ are two
valid states. Because the qplex $\cP$ is a maximal consistent set, the
convex combination
\begin{equation}
  p_\lambda = \lambda p_1 + (1-\lambda) p_2
\end{equation}
must also belong to $\cP$. The reflection principle gives this state
meaning: It is my belief now about what a reference measurement might
yield in the future, provided that I believe now that some
intermediate action will lead me to update my gambling commitment
either to $p_1$ or to $p_2$ with probabilites $\lambda$ and $1 -
\lambda$ respectively. Define
\begin{equation}
  q_1(E) = \mu_E(p_1),\ q_2(E) = \mu_E(p_2),\ q_\lambda(E) = \mu_E(p_\lambda).
\end{equation}
If I believe now that my future gambling commitment about the event
$E$ will be either $q_1(E)$ or $q_2(E)$, then my van Fraassen
reflection probability for $E$ is
\begin{equation}
  q_\lambda(E) = \lambda q_1(E) + (1-\lambda)q_2(E).
\end{equation}
We begin to see why, morally, $\mu_E$ must be a linear function on
the qplex $\cP$.  Specifically, if I am willing to treat $p_\lambda$
just like any other probability vector for the reference measurement,
even though I know I calculated it by taking a weighted average, then
I can say
\begin{equation}
  q_\lambda(E) = \mu_E(p_\lambda),
\end{equation}
and equating our two ways of writing $q_\lambda(E)$ shows that $\mu_E$
must preserve convex combinations:
\begin{equation}
  \mu_E\left(\sum_k \lambda_k p_k\right)
  = \sum_k \lambda_k \mu_E(p_k),
  \ \sum_k \lambda_k = 1.
\end{equation}
This means that $\mu_E$ must be an affine function, that is, a
linear function of~$p$ possibly combined with a constant offset.
Suppose we could make that constant vanish.  Then $\mu_E$ would be a
linear, nonnegative function on the qplex.  But we know what all such
functions look like: We can write each one as an inner product of~$p$
with some vector, and we have the full set of vectors whose inner
product with $p$ is nonnegative for all $p \in \cP$, thanks to
self-polarity. Thus, we would have
\begin{equation}
  \mu_E(p) = \kappa_E \inprod{s_E}{\Phi p},
\end{equation}
for some constant $\kappa_E$ and some vector $s_E \in \cP$.

We know that the offset has to vanish at least some of the time,
because the reference measurement is itself a valid experiment.
``Computing'' the probabilities for the outcomes of the reference
measurement in terms of themselves amounts to saying $\mu_E(p) = p(i)$
for some $i = 1,\ldots,N$. In other words, $\mu_E$ just plucks out a
particular element of~$p$.

We aren't actually \emph{obligated} by coherence to treat a van
Fraassen reflection state in exactly the same way we would treat the
same state arrived at by another method. Alice may believe now that
her belief tomorrow will be $P_k$ for $k = 1$ or 2. Alice also
believes now that she will have had the experience which will lead her
to select her new gambling commitment. In other words, Alice believes
that she will have an outcome $k$ in hand. That is extra information
which could affect what functions she is willing to apply to her
matrix of joint probabilities.  The crucial fact is that she is
willing \emph{sometimes} to treat a van Fraassen reflection state like
any other. Sometimes, she will believe that she will have experienced
a value of~$k$, and that the consequences will be restrained. Modest,
one might say. In practice, this seems to mean treating the event $k$
as affecting one's choice of $p$ or of $r$ but not both.

The application of van Fraassen's reflection principle to
\emph{conditional} probabilities is a surprisingly subtle issue. I
continue to hold out hope for the opportunity to coauthor a technical
paper on the topic, but in the meantime, we can proceed as follows.

We consider reference measurements having $N$ possible outcomes (and
so $p \in \Delta \subset \bbR^N$) and measurements on the ground
having $M$ possible outcomes (meaning that $q \in \bbR^M$).  Applying
the reflection principle separately to $p$ and to $r$, in the manner
outlined above, we find that
\begin{equation}
  q = A_M(r,p)
\end{equation}
where $A$ is a biaffine map. That is, when we fix either $r$ or $p$,
then $A_M$ becomes a function of the other argument that combines a
linear transformation with an offset.  Then by standard
results~\cite{MacLane:1999} we have the decomposition
\begin{equation}
  A_M(r,p) = L_{M,0}(r,p) + L_{M,1}(r) + L_{M,2}(p) + C_M.
\end{equation}
Here, $L_{M,0}$ is a bilinear map, both $L_{M,1}$ and $L_{M,2}$ are linear
functions, and $C_M$ is a constant vector.

From the basic setup of the ground-versus-sky picture, we know that
the set of valid measurements must contain $r_F$, the standard or
reference measurement itself.  When $M = N$ and $r = r_F$, we have
\begin{equation}
  q = A_N(r_F,p) = p.
\end{equation}
Moreover, regardless of $M$, we know that when an agent contemplates
following the sky path, they expect to assign one of the states
$\{e_k\}$ after performing the reference measurement. By definition,
$r(j|k)$ is the probability of obtaining the $j$th outcome of the
measurement $r$ given that the ``preparation'' is the vector $e_k$.
Therefore, if we fix $p = e_k$, then
\begin{equation}
  A_M(r,e_k) = L_{M,0}(r,e_k) + L_{M,1}(r) + L_{M,2}(e_k) + C_M
\end{equation}
must be linear in $r$.  Everything that is not linear in $r$ on the
right-hand side must vanish, meaning that
\begin{equation}
  L_{M,2}(e_k) = -C_M.
\end{equation}

But any properly normalized $p$ is a linear combination of the
$\{e_k\}$ with coefficients (possibly negative) that sum to~1, because
the inverse of a stochastic matrix is quasistochastic.  Therefore,
\begin{equation}
  L_{M,2}(p) = L_{M,1}(r_F)\ \forall\ p \in \Delta.
\end{equation}
Now, we only need to know $L_{M,2}$ on the probability simplex, since that
is where the physically meaningful inputs live, and so we have reduced
our original biaffine form to
\begin{equation}
  A_M(r,p) = L_{M,0}(r,p) + L_{M,1}(r).
  \label{eq:biaffine-reduced}
\end{equation}

When $r = r_F$, the above expression must reduce to $p$:
\begin{equation}
  L_{N,0}(r_F,p) + L_{N,1}(r_F) = p.
\end{equation}
This is satisfied by taking $L_{N,0}(r_F,p)$ to be the identity
transformation on~$p$ and $L_{N,1}(r_F)$ to be the zero
vector.

Thus, the reflection principle constrains \emph{the possible forms of
  ``contextuality''}, by restricting how probabilities on the ground
can depend upon the entirety of $r$, instead of individual rows within
it. For the time being, we will assume that $L_{M,1}$ will vanish in
general. It is possible that a more detailed development of
conditional reflection will allow us to narrow the functional form of
$A_M$ further, and it is also conceivable that a sufficiently careful
statement of the ``zeroth law of quantum mechanics'', which specifies
when scenarios are equivalent, could cut off awkward types of
context-dependence at the outset.

Probability theory is a very linear theory. In this project, we are
trying to squeeze the difference between quantum physics and classical
intuition into as small a corner as possible. Thus, when we introduce
the idea of a reference measurement, we want the resulting theory to
inherit as many of probability's natural linearities and affinities as
are feasible. To do otherwise would be to build a pathological theory,
and that's not what we're interested in doing. We have arrived at a
quantitative statement, that the best replacement for the classical
Law of Total Probability is also a bilinear map, from the more
conceptual consideration that a belief about beliefs can be treated as
a belief itself. Drilling futher down into this topic may require a
deeper investigation into why and how probability theory itself is a
viable theory for managing preferences, expectations and utilities.

\begin{comment}
The next step is a little unusual: We reflect on \emph{conditional}
probabilities. Imagine that we hold the ``preparation'' fixed and
consider two possible measurements with the same number of
outcomes. Then reflection implies convex-linearity for each value of
the index $j$:
\begin{equation}
  r_\lambda(j|i) = \lambda r_1(j|i) + (1-\lambda) r_2(j|i).
\end{equation}
\end{comment}
The rows of any given conditional probability matrix $r$, which are
labeled by the index $j$, won't necessarily be normalized nicely, so
let's write them as
\begin{equation}
  r(j|i) = N\gamma_j s_j(i),
\end{equation}
where $s_j$ is a probability vector. The way we've chosen for writing
the prefactor will turn out to be convenient later. Linearity on both
inputs means that
\begin{equation}
  q(j) = N\gamma_j s_j^{\rm T} A\, p
\end{equation}
for some matrix $A$. We identify $A$ with the $\Phi$ we defined in
Eq.~(\ref{eq:gen-Phi}) by identifying the basis distributions as the
post-measurement states of the reference experiment. When we feed in a
basis distribution $e_k$, we are setting $p$ equal to a column of
$\Phi^{-1}$, and so our result for $q(j)$ just reads out
$r(j|k)$. Self-polarity then implies that the set of valid $s_j$ is
the same as the set of valid $p$, that is to say, the qplex $\cP$.

We have arrived at the \emph{generalized urgleichung}:
\begin{equation}
  q = r\, \Phi\, p.
\label{eq:gen-urgleichung}
\end{equation}

The matrix $\Phi$ preserves the flat probability distribution
$c$. Consequently, we have an interpretation for $\gamma_j$: It is
just the probability for obtaining the $j$\textsuperscript{th} outcome
of the measurement $r$, given the state of complete indifference for
the reference experiment.

The reflection principle has led us to linearity, and maximality has
brought us to the fact that a qplex must be self-polar, which in turn
allowed us to conclude that the set of valid measurement matrices, $\cR$,
is built from the set of valid preparations, $\cP$. In other words, our
mental model of a measurement is a set of preparations, possibly
weighted. To explore this in more conceptual detail, we echo the
discussion from~\cite{Appleby:2017} with minor edits:
\begin{quotation}
\noindent What conditions must an object meet in order to qualify as a
piece of laboratory apparatus?  Classically, a bare minimum
requirement is that the object has a set of distinguishable
configurations in which it can exist.  These might be positions of a
pointer needle, heights of a mercury column, patterns of glowing
lights and so forth.  The essential point is that the system can be in
different configurations at different times:\ A thermometer that always
reports the same temperature is useless.  We can label these
distinguishable configurations by an index $j$.  The
\emph{calibration} process for a laboratory instrument is a procedure
by which a scientist assigns conditional probabilities $\{r(j|i)\}$ to the
instrument, relating the readout states $j$ to the inputs $i$.  In
order to make progress, we habitually assume that nature is not so
perverse that the results of the calibration phase become completely
irrelevant when we proceed to the next step and apply the instrument
to new systems of unknown character.

But what if nature \emph{is} perverse?  Not enough so to forbid the
possibility of science, but enough to make life interesting.
Quantitatively speaking, what if we must modify the everyday
assumption that one can carry the results of a calibration process
unchanged from one experimental context to another?

\emph{The urgleichung is just such a modification.}  The conditional
probabilities $\{r(j|i)\}$ do not become irrelevant when we move from
the upper path in Figure~\ref{figure1} to the lower, but we do have to
use them in a different way.

In quantum physics, we no longer treat ``measurement'' as a passive
reading-off of a specified, pre-existing physical quantity.  However,
we do still have a counterpart for our classical notion of a system
that can qualify as a laboratory apparatus.  Instead of asking whether
the system can exist in one of multiple possible classical states, we
ask whether our overall mesh of beliefs allows us to consistently
assign any one of multiple possible catalogues of expectations.  That
is, if an agent Alice wishes to use a system as a laboratory
apparatus, she must be able to say now that she can conceive of
ascribing any one of several states to it at a later time.

The analogue of classical uncertainty about where a pointer might be
pointing is the convex combination of the states $\{s_j\}$.
Therefore, our basic mental model of a laboratory apparatus is a
polytope in~$\cP$, with the $\{s_j\}$ as its vertices.  The conclusion
that we build $\cR$ from $\cP$ says that \emph{Alice can pick up any
  such apparatus and use it as a ``prosthetic hand'' to enrich her
  experience of asking questions of nature.}
\end{quotation}

Instantly updating to a post-measurement state on the out-sphere of a
qplex is like converting a massive body into light: We leap to an edge
that more mundane transformations could not attain. \emph{Maximal
  certainty} is what plays the role of the speed of light or absolute
zero. Our third law, which tells us that reversible transformations
can asymptotically approach maximal certainty, where the
post-measurement states of the reference experiment live, brings us to
the next stage of the story.

Marcus Appleby made a new branch of mathematics available to us by
defining what I will call a \emph{qplectic cone
  theory}~\cite{Appleby:unpublished}. First, we need the notion of a
\emph{convex theory}, which is a triple $(V, \cone, \orderunit)$,
where $V$ is a real $d^2$-dimensional vector space whose inner product
we write $\varinprod{\cdot}{\cdot}$. The cone $\cone \subset V$ is
self-dual with respect to this inner product, and $\orderunit$ is an
element in the interior of $\cone$ that we'll call the \emph{order
  unit}. We use the order unit to define the hyperplane of normalized
elements:
\begin{equation}
  \cH = \{v \in V: \varinprod{\orderunit}{v} = 1\}.
\end{equation}
The intersection of this hyperplane with the cone $\cone$ is the
\emph{state space}:
\begin{equation}
  \Omega = \cH \cap \cone.
\end{equation}
A qplectic cone theory is a convex theory $(V,\cone,\orderunit,
\{e_j\})$ where the state space $\Omega$ contains a regular simplex
$\{e_j: j=1, \ldots, d^2\}$ such that
\begin{equation}
  \varinprod{e_j}{e_k} = \frac{d\delta_{jk}+1}{d+1}.
\end{equation}
As a minor technical point, the inner product and order unit in a
qplectic cone theory are scaled such that the supremum of $|p|^2$ over
all $p\in\Omega$ is unity.

The elements $\{e_j\}$ can be shown to form a basis, with a dual basis
\begin{equation}
  \bar{e}_j = \frac{d+1}{d}e_j - c,
\end{equation}
where we have defined the center point
\begin{equation}
  c = \frac{\orderunit}{\varinprod{\orderunit}{\orderunit}}
  = \frac{1}{d}\orderunit.
\end{equation}
The dual basis is a resolution of the order unit:
\begin{equation}
  \sum_k \bar{e}_k = \orderunit \,;
\end{equation}
as is the original basis, up to scaling:
\begin{equation}
  \sum_k e_k = d\orderunit.
\end{equation}

The key result is the following. Let $(V,\cone,\orderunit, \{e_j\})$
be a qplectic cone theory. Define $f:V \to \bbR^{d^2}$ to be the
affine bijection
\begin{equation}
  f(p) = (p_1,\ldots,p_{d^2})^{\rm T},
  \ p_j = \frac{d\varinprod{\bar{e}_j}{p} + 1}{d(d+1)}.
\end{equation}
Then $f(\Omega)$ is a qplex. Conversely, given an arbitrary qplex, we
can realize it as the image under the appropriate affine bijection of
the state space $\Omega$ of some qplectic cone. This correspondence
carries over to measurements, which in a qplectic cone theory are
resolutions of the order unit $\orderunit$ into elements of the cone
$\cone$.

John DeBrota has worked out the details for generalized qplexes, where
the dimension and the bounds are not fixed to the values familiar from
quantum theory. In this broader context, our convex theory will again
be a tuple $(V, C, \orderunit, \{e_i\})$, but with
\begin{equation}
\varinprod{e_i}{e_j} = 1 + NL(\delta_{ij} - 1).
\end{equation}
The barycenter $c$ is given by
\begin{equation}
  c = \frac{1}{N}\sum_k e_k
  = \frac{\orderunit}{\varinprod{\orderunit}{\orderunit}}.
\end{equation}
The scaling of the $\orderunit$ is
\begin{equation}
\varinprod{\orderunit}{\orderunit} = \frac{1}{1 + L - NL}.
\end{equation}
The dual basis to $\{e_i\}$ is given by the points
\begin{equation}
\bar{e}_i = \frac{1}{NL}\left(e_i - \frac{1-NL}{1+L-NL} c\right).
\end{equation}
It follows that the dual basis is again a resolution of the order
unit.  As before, we take a generalized qplex to be a set $\cP \subset
\Delta$ such that for all $p,p' \in \cP$, we have
\begin{equation}
L \leq \inprod{p}{p'} \leq U,
\end{equation}
and $U$ is given by the length of the basis distributions:
\begin{equation}
U = 1 + L(N-1)(NL-2).
\end{equation}
The vertices of the basis simplex have coordinates
\begin{equation}
e_k(j) = (1-NL)\delta_{jk} + L.
\end{equation}
A bijection is established by $f: V \to \mathbb{R}^N$, where
\begin{equation}
f(v) = (p_1,\ldots,p_N)^{\rm T},
\ p_i = (1-NL)\varinprod{\bar{e}_i}{v} + L.
\end{equation}
For any $v \in \Omega$,
\begin{equation}
v = \sum_j \frac{p_j - L}{1-NL} e_j.
\end{equation}

Convex cones can be classified once we postulate a sufficient degree
of symmetry. It is known that convex cones which are
\emph{homogeneous} are all isomorphic to algebraic structures of a
genre that, in turn, has been classified exhaustively. Homogeneous
cones are those for which any point in the interior of the cone can be
mapped to any other by some isomorphism of the cone. The
classification task is achieved by the Koecher--Vinberg theorem, which
establishes that there exists a one-to-one correspondence between
these highly symmetric cones and ``formally real Jordan
algebras''~\cite{Faraut:1994, Wilce:2016}. The Koecher--Vinberg
theorem is one of the two mathematical results upon which this
approach to reconstructing quantum theory relies; the other, more
empirical in character, is that SICs appear to be a quintessentially
complex-vector-space phenomenon.

It took me a long time to become accepting of homogeneity as a
candidate axiom. Originally, it seemed like the kind of constraint
that the mathematicians impose just so they have a set they can
classify. Furthermore, the objects being classified did not seem
particularly ``natural'' or well-motivated from our starting point. My
first temptation to reconsider was Marcus Appleby's construction of
what I have termed qplectic cones~\cite{Appleby:unpublished}.  Later
still, building upon this, I realized that ``doing as the
mathematicians do'' might be exactly what the project called
for. Maximal symmetry and maximal abstraction can be, at the right
juncture, an expression that the physical content lives at a different
level.

Prior work has noted that given a qplex, the urgleichung is the
natural way to calculate probabilities for other experiments in terms
of the reference measurement, and in consequence, ``posteriors from
maximal ignorance are valid priors''~\cite{Appleby:2017}. Homogeneity
is an extension of this idea. Not a trivial one, as the existence of
highly asymmetrical qplexes will attest, but an amiable one.  Instead
of merely noting that any point $p$ in our qplex $\cP$ is a valid
post-measurement state for some experiment given the flat vector $c$
as input, we posit an update rule, a map from the qplex to itself that
takes $c$ to $p$ (up to normalization). One justification for this
elaboration comes from the convexity of the qplex. Because the flat
vector $c$ can be decomposed as a convex combination of other states
(in many ways), we can regard $c$ as a van Fraassen reflection state,
an ``expected future expectation'', and a valid transformation of $c$
ought to be a valid transformation of the points into which we
decompose it as well. (If an event leads me to update my state of
belief to $p$ if I currently believe $c$, what does it prompt me to do
if I believe $s$? Because $s$ and $p$ are taken from the qplex $\cP$,
this sets us on the path to defining a binary operation on~$\cP$.)
Homogeneity is motivated by asking that these transformations be
nontrivial and reversible whenever possible.

Another way to express this motivation is to note that we are trying
to put an algebraic structure on the qplex, so we should consider
where the notion of an \emph{algebra} came from.  As students,
somewhere in our education, we learned to think of
\emph{multiplication} as \emph{scaling}:\ The quantity ``$a$ times
$b$'' is where $b$ lands when we apply the transformation of space
that sends 1 to~$a$. (Usually, we might first meet this picture in the
setting of complex numbers, though not always~\cite{Gonick:2015}.)  If
we have an isomorphism of the qplectic cone that sends the order unit
$\orderunit$ to a point $a$, we can define the product $a \star b$ by
finding where this transformation takes the point $b$. This is how we
``treat points as numbers'', i.e., how we give the qplectic cone an
algebraic structure.

A final justification for homogeneity of the qplectic cone is that
homogeneity is a property that holds classically~\cite{Barnum:2019},
and we are doing our best to localize the conceptual departure from
classicality in a single place, the principle of quantum vitality. In
classical probability theory, we can use the all-ones vector as our
order unit, and obtain any point in the interior of the probability
simplex trivially, by multiplying each entry by $p(i)$. The quantum
analogue of this map is slightly more elaborate:\ The operation
\begin{equation}
  L_\rho(X) = \rho^{1/2} X \rho^{1/2}
\end{equation}
sends the identity matrix to $\rho$ and is invertible whenever $\rho$
is not singular.

In our efforts to reconstruct the quantum formalism, we have focused
on the concept that nonclassicality is revealed by the \emph{meshing}
of expectations for \emph{different} measurements. If you only
consider a single experiment, you can just write a probability
distribution over its outcomes. Rolling this concept around a while
suggests the idea that for any given state inside a generalized qplex,
there should be at least one measurement with respect to which that
state ``looks classical''. More specifically, it motivates carrying
over a property from classical probability theory. Any point in a
probability simplex is automatically a convex combination of
completely distinguishable distributions. From the conditions defining
a generalized qplex, we can deduce a bound on the maximum size of a
mutually maximally distant set, which above we denoted $d$. So, let's
say that any point inside a generalized qplex can be written as a
convex combination of at most $d$ MMD pure points. Moreover, we insist
that any set of less than $d$ MMD states can always be completed to a
full set of $d$. This ensures that their convex hull contains the flat
probability vector. If there is some point for which this is not true,
then there is a preparation which never looks quasiclassical with
respect to any experiment; this might not explicitly contradict our
basic assumptions, but it does seem unnatural.

We have here a justification for our generalized qplexes to be
\emph{spectral}~\cite{Wilce:2016}.

Going a little further: In order for a subtheory to be valid, the
transformations in the subtheory must be respected by the full theory
into which it embeds. So, if we really want an MMD set carved out from
a generalized qplex to define a subtheory, the allowed transformations
of the generalized qplex have to let us do to the simplex given by the
MMD set anything that we could do to a classical probability simplex
of the same dimension. But a classical probability simplex is
homogeneous, or rather, it's a slice through a homogeneous cone. One
can go inward to the center and back out again via isomorphisms of the
cone. We know that the barycenter of an MMD set is the flat
probability vector. Consequently, we can take any nonsingular state in
to the center and back out again.  Homogeneity of the full qplectic
cone follows from homogeneity of the subtheories.

Now that we are at least content enough with the idea of putting an
algebra on a qplex to jump in and see where it goes, we can
investigate what kinds of algebraic structure are consistent with the
features that are common to all qplexes. In
section~\ref{sec:hilbert-qplex}, we will touch upon the possibility of
an alternative path to establishing an algebra.

\begin{comment}
We now invoke the first and second laws and consider two states
$p_1,p_2 \in \cP$, each of which implies certainty for some
experiment, and which are maximally distinguishable. We have
\begin{equation}
  \inprod{p_1}{p_2} = L,
\end{equation}
and there must exist some $\mu_E$ that yields probability 1 when given
the input $p_1$ and probability 0 when given the input $p_2$. For any
$p \in \cP$, we must have $\mu_E(p) \in [0,1]$, and moreover,
\begin{equation}
  \mu_E(\lambda p_1 + (1-\lambda)p_2) = \lambda.
\end{equation}
Because $\mu_E$ is a linear function, we can write it as an inner
product:
\begin{equation}
  \mu_E(p) = \inprod{\mu_E}{p}.
\end{equation}

Note that we can rewrite the fundamental inequalities as
\begin{equation}
  0 \leq \frac{\inprod{p_1}{p_2} - L}{U - L} \leq 1.
\end{equation}
We can conveniently package this as a bilinear form,
\begin{equation}
  0 \leq p_1^{\rm T} M p_2 \leq 1,
\end{equation}
if we define the matrix
\begin{equation}
  M = \frac{1}{U-L} I - \frac{L}{U-L}J,
\end{equation}
using $J$ to stand for the all-ones matrix (the Hadamard identity).
The matrix $M$ is easily inverted, yielding
\begin{equation}
  M^{-1} = (U-L)I + L \frac{U-L}{1-NL} J.
\end{equation}
\end{comment}

\section{Narrowing the Choice of Algebras}

A Jordan algebra is an abstraction of some of the properties enjoyed
by observables in quantum theory~\cite{Barnum:2019, Mccrimmon:1978}.
In what follows, we will restrict ourselves to considering
finite-dimensional algebras only.  To build a Jordan algebra, we need
a set of quantities that we can combine with a product operation, and
this product must satisfy two conditions.  First, it needs to be
commutative:
\begin{equation}
x \circ y = y \circ x.
\end{equation}
And second, rather than being so kind as to satisfy the associative
law, the algebra satisfies the \emph{Jordan identity,}
\begin{equation}
(x \circ y) \circ (x \circ x)
 = x \circ (y \circ (x \circ x)).
\end{equation}
One way to build a Jordan algebra is to take a matrix algebra and
define the Jordan product in terms of a symmetrized combination of
matrix products:
\begin{equation}
x \circ y = \frac{1}{2}(xy + yx).
\label{eq:anticomm}
\end{equation}
A Jordan algebra is \emph{formally real} if a sum of squares only
vanishes when each individual element being squared vanishes:
\begin{equation}
x_1^2 + x_2^2 + \cdots + x_n^2 = 0
\ \Rightarrow\ x_1 = x_2 = \ldots = x_n = 0.
\end{equation}
An \emph{ideal} in a Jordan algebra is a subspace that ``absorbs
multiplication''.  A subspace $B \subseteq A$ is an ideal of~$A$ if,
for every $b \in B$, we have $x \circ b \in B$ for all $x \in A$.  We
call a Jordan algebra \emph{simple} if its only ideals are the empty
set and the whole algebra itself.  Any formally real Jordan algebra
can be written as a direct sum of simple Jordan algebras.  Thus,
classifying all formally real Jordan algebras boils down to the task
of classifying the simple ones.  Jordan, von Neumann and Wigner did
this, back in the day.  There are three infinite sequences, defined by
applying Eq.~(\ref{eq:anticomm}) to matrix algebras of $d \times d$
self-adjoint matrices built from the real numbers, the complex numbers
and the quaternions respectively.  In addition, there is another
infinite family built from the direct sum of vector spaces
$\mathbb{R}^d \oplus \mathbb{R}$.  An element of such an algebra is an
ordered pair
\begin{equation}
(v,\alpha), \hbox{ with } v \in \mathbb{R}^d,\ \alpha \in \mathbb{R}.
\end{equation}
The product rule for this algebra is
\begin{equation}
(v,\alpha) \circ (w,\beta) = (\alpha w + \beta v,
 v \cdot w + \alpha\beta).
\end{equation}
This is the family of ``spin factors''.  Finally, there is an
exceptional case, the \emph{Albert algebra,} defined by applying
Eq.~(\ref{eq:anticomm}) to $3 \times 3$ self-adjoint matrices of
octonions.

We will see in the next section that we can parameterize generalized
qplex theories by relating the cardinality of the reference
measurement, $N$, to the maximum size of an MMD set:
\begin{equation}
  N = d + \qbar \frac{d(d-1)}{2},
\end{equation}
where $\qbar$ is a nonnegative integer. The case $\qbar = 0$ collapses
to classical probability theory: The qplex is just the probability
simplex in $\bbR^d$, the lower bound $L$ drops to zero, and the MMD
states are Kronecker delta functions. When we set $\qbar = 2$, we
recover the quantum formula, $N = d^2$, while $\qbar = 1$ and $\qbar =
4$ reproduce the scaling relations of the real and quaternionic ``foil
theories'' to quantum mechanics. The spin-factor theories have state
spaces that are Euclidean balls, and so they can only exist for $d =
2$: Given any extremal point, there is exactly one antipodal point.

In a Jordan theory, the space of observables forms a simple Jordan
algebra.  The state space is the set of positive observables that have
trace equal to~1.  This generalizes quantum theory:  In the quantum
theory of a finite-dimensional system, the observables are
self-adjoint complex matrices, and the states are the density
operators, which are positive semidefinite matrices with unit trace.
Generally, then, in a Jordan theory, any state is also an observable.
We can also define effect operators in the analogous manner to quantum
theory, and then we can construct the Jordan-theoretic version of
POVMs, as collections of effects which sum to the identity.

We can see the general plan for the final step of a Jordan-theoretic
reconstruction of quantum theory.  One way or another, we wish to
select the \emph{complex} option out from among the real, the
quaternionic, the octonionic and the spin-factorial.  (In what
follows, we will be expanding upon a suggestion made in~\cite[\S
  7]{Fuchs:2019}.)

The constant $N$ is fixed by the number of real parameters needed to
specify a density operator.  If we cannot find $N$ equiangular pure
states within the state space of a Jordan theory, then we cannot
satisfy the generalized urgleichung, Eq.~(\ref{eq:gen-urgleichung}).
Conversely, if we assert that the generalized urgleichung is valid for
all dimensions, then we rule out all classes of Jordan theories which
do not admit a full set of equiangular pure states.

Why should we focus on \emph{simple} algebras? Because an algebra
splitting into a direct sum, $\cA = \cA_1 \oplus \cA_2$, represents
\emph{classical ignorance}. This models the situation where our system
of interest might be one for which we use the algebra $\cA_1$, or it
might be one for which we use the algebra $\cA_2$, and the choice
between them is a classical bit. Recall that we admitted the
possibility of there being many possible reference measurements, of
which the most conceptually illuminating --- the analogues of Carnot
engines or light clocks --- are those for which perfect
distinguishability corresponds to the constant overlap $L$. If the
algebra of the qplectic cone factored, we could make a reference
measurement whose elements are direct sums of operators in the
separate factors, and states from different factors would have
nonoverlapping probabilistic representations. The requirement that all
reference measurements avoid this ensures that the algebra has just
one factor.  Another heuristic is that, if the basis distributions
$\{e_k\}$ map to algebra elements that are spread across multiple
factors, then they represent both classical ignorance (the choice of
factor) and quantum uncertainty, and so they would not be states of
maximal confidence, which we know they are. So, the basis
distributions must each map to a state localized within a factor, and
since by construction they cannot be orthogonal states, they must all
lie within the same factor, along with the entirety of the state space
they span.\footnote{Recall how Barnum, M\"uller and
  Ududec~\cite{Barnum:2014} obtain simplicity: by noting that there
  can't be two different kinds of orthogonality.  Two orthogonal pure
  states from the same factor define a Bloch ball (possibly not
  three-dimensional), while two pure states from different factors
  define a quasi-classical bit.  Having two qualitatively different
  kinds of faces in this way is ruled out by their first two
  postulates (``classical decomposability'' and ``strong symmetry'').}

While much remains unknown about the subject of equiangular lines, all
indications to date are that the complex option --- that is, ordinary
quantum theory --- is the only possibility consistent with this
requirement.  For the real case, we would require
\begin{equation}
N_{\bbR} = \binom{d+1}{2}
\end{equation}
equiangular lines for each value of the dimension $d$.  This is an
upper bound that one can deduce from linear algebra, but not every
dimension can sustain a full set of $N_\bbR$ equiangular
lines~\cite{VanLint:1966, Lemmens:1973, Delsarte:1975, Seidel:1983,
  Sloane:2015, Greaves:2016}.  In fact, the only known dimensions in
which the bound is attainable are $d = 2$, 3, 7 and 23.  One can prove
that the only possibilities for having a full set are when $d$ is 2, 3
or a value 2 less than the square of an odd number. (Such a
mathematician's statement!) The conjecture that the bound can be
attained whenver $d + 2$ is an odd square was disproved by an explicit
counterexample for dimension 47, back in 2002~\cite{Greaves:2016}. So,
the real-vector-space analogue of quantum theory is ruled out: We
cannot even formulate the generalized urgleichung in the EPR-type
case, $d = 4$; nor in the GHZ-type scenario, $d = 8$. By contrast, not
only do we have SICs in $\bbC^4$ and $\bbC^8$, but they even have
rather remarkable properties above and beyond the symmetry implied by
the definition~\cite{Bengtsson:2017, Zhu:2010, Szymusiak:2016,
  Stacey:2016b, Andersson:2019}.

The quaternionic case is much less understood than either the real or
the complex, but searches to date have failed to find a quaternionic
SIC in any $d > 3$~\cite{Cohn:2016}.  That is, we require for all
dimensions $d$ a set of
\begin{equation}
N_{\bbH} = d(2d-1)
\end{equation}
equiangular projection operators for a symmetric, informationally
complete quaternionic measurement~\cite{Graydon:2011}. Corroborating
the work reported in the literature~\cite{Cohn:2016}, Matt Weiss has
conducted numerical searches in $d = 4$, finding that one can have as
many as 21 equiangular lines, but not the full 28 that are necessary
to saturate the Gerzon bound and make a SIC in
$\bbH^4$~\cite{Weiss:2022}. Granting that these measurements are
probably not available in all dimensions, we rule out the quaternionic
analogue of quantum theory. Moreover, there are arguments that the
real and quaternionic theories should be considered together or not at
all~\cite{Barnum:2016}, and since we \emph{know} that ``real SICs'' do
not exist for all dimensions, we can set both $\bbR$ and $\bbH$ aside.

There \emph{does} appear to be an ``octonionic SIC'' in $d = 3$,
corresponding to the Albert algebra. Cohn, Kumar and
Minton give an existence proof and a numerical solution, though not an
exact construction~\cite{Cohn:2016}. However, we can say confidently
that we do not want our theory to stop at three-level systems; this
case may be more interesting for group theory than for quantum
physics~\cite{Wilson:2009}. For example, it may be possible to think
of each point in the Leech lattice as a ``Hamiltonian'' for an
``octonionic qutrit'', since the Leech lattice can be embedded in the
space of traceless self-adjoint $3 \times 3$ matrices over the
octonions~\cite{Baez:2014}. The symmetry group of the generalized
qplex for this unusual system would be what the specialists call
$\mathrm{F}_4$, and the symmetries of the set of Leech
``Hamiltonians'' would give the Conway groups. A similar claim could
indeed be made for the equiangular lines in~$\bbR^d$, since one
construction of the maximal set in $\bbR^7$ extracts it from the
$\mathrm{E}_8$ lattice, and the maximal set in $\bbR^{23}$ is derived
from the Leech lattice, thereby connecting the ``real SICs'' with many
other optimality problems~\cite{Viazovska:2017, Cohn:2019}.

The real and octonionic ``foil theories'' may well be having a subtle
liason with the Leech lattice, but for physics purposes, it appears
that the complex option is the only theory left standing. We can
articulate the lesson as follows:

Quantum theory is the maximally symmetric probabilistic theory that
embraces vitality, and we pursue maximum symmetry because we want to
present that vitality in the clearest manner possible.

Our route to solidifying this lesson requires two plausible but
unproven conjectures: first, that SICs always exist in~$\bbC^d$, and
second, that a SIC fails to exist for some $\bbH^d$. The former
conjecture has been shown to follow in turn from plausible, but
sophisticated, conjectures in algebraic number theory~\cite{Appleby2025}.

\section{Interlude: Pause for a Milkshake}
\begin{quotation}
\noindent One summer in graduate school I was a student of [Edward]
Condon's. I remember vividly his account of being brought up before
some loyalty review board:

``Dr.\ Condon, it says here that you have been at the forefront of a
revolutionary movement in physics called'' --- and here the inquisitor
read the words slowly and carefully --- ``quantum mechanics. It
strikes this hearing that if you could be at the forefront of one
revolutionary movement \ldots\ you could be at the forefront of
another.''
\begin{flushright}--- Carl Sagan, \booktitle{The Demon-Haunted
    World}~\cite{Sagan:1995}
\end{flushright}
\end{quotation}

Physics has, most likely, as morally checkered a past as most modern
vocations. We can without contradiction observe simultaneously that
physics suffered in the McCarthy era --- with Melba Phillips being
blacklisted for five years, for example~\cite{PT:2019} --- and that
the profession was in essence the long-term planning department of the
military-industrial complex. It is in this context that we pause to
note that Pascual Jordan was a morally repugnant human being. After
publishing extremist political screeds in the 1920s under a pseudonym,
he abandoned pretence in 1933, joining the Nazi Party and signing up
to be a brownshirt. He tried to defend quantum mechanics from the
charge of being ``Jewish physics'' \ldots\ because he said it was an
antidote to Bolshevik materialism. Perhaps many organizations can be
changed from within, as Jordan apparently tried to justify himself to
Bohr after the war~\cite{Bernstein:2005, Beyerchen:2018}, but we doubt
that any fascist parties are among them.

Now I know how specialists in Teichm\"uller spaces feel, if they have
any conscience. (I recall that at one point, Oswald Teichm\"uller's
Wikipedia page described him bluntly as ``a mathematician and Nazi''.)

But there is a flipside, because legacies are funny things. To
paraphrase Horace, many brave and bloody men lived before Agamemnon,
but they all sank into darkness for lack of a poet. Macbeth ruled
Scotland for 17 years, but he is now forever the king who trafficked
with witches who were the Fates. In that vein, less
grandiosely:\ Someone out there might be learning about Jordan
algebras for the first time because of this, an essay by a nonbinary
physicist. And Jordan wouldn't be too happy about that.

%\pagebreak
\section{More Details Regarding Reconstruction}
\begin{flushright}
  My sins my own, they belong to me. Me! \\
  People say ``beware!'' But I don't care. \\
  The words are just rules and regulations \\to me. Me! \\
  \smallskip
--- Patti Smith
\end{flushright} 

It is worth investigating a little more deeply the idea that the
inequality $\inprod{p_1}{p_2} \geq L$ is a statement of quantum
vitality. We can do so thanks to a variant of Lemma 4 in the ``MIC
Facts'' survey~\cite{DeBrota:2018b}. A \emph{minimal informationally
  complete} measurement, or \emph{MIC}, is what we get when we relax
the requirement that our reference experiment be symmetric: It is a
set of positive semidefinite operators $\{E_i:i=1,\ldots,d^2\}$ that
sum to the identity matrix. This furnishes a reference measurement for
quantum states ascribed to $d$-dimensional systems. Specifically, we
can express any quantum state $\rho$ as an expansion
\begin{equation}
  \rho = \sum_i p(E_i) \widetilde{E}_i,
\end{equation}
where $p(E_i)$ is the Born-rule probability $\tr(\rho E_i)$ and
$\{\widetilde{E}_i\}$ is the dual basis of the MIC.

Assume for the moment that all the MIC elements $\{E_i\}$ are
proportional to rank-1 projectors. Can two orthogonal states $\rho$
and $\sigma$ have orthogonal probabilistic representations $p_\rho$
and $p_\sigma$? Not for qubits, that's for sure: When $d = 2$, we can
have at most \emph{one} zero in a probability vector. At most one of
the $\{E_i\}$ can be orthogonal to a given vector, and by linear
independence, two different MIC elements cannot be proportional to
each other. What about $d > 2$? Let $\ket{\psi}$ and $\ket{\phi}$ be
two orthogonal states. Together they define a qubit-sized
subspace. Now, project the MIC into this subspace. If all the images of the
MIC elements lined up with $\ket{\psi}$ or $\ket{\phi}$, then the MIC
would not be informationally complete on that subspace, because
spanning requires four elements. Therefore, at least one of the
$\{E_i\}$ must be nonorthogonal to both $\ket{\phi}$ and
$\ket{\psi}$. The slot in the probability distributions corresponding
to this MIC element must be nonzero for both. Consequently, the
Euclidean inner product of these vectors cannot be zero.

More generally, we can drop the rank-1 condition. If $\rho$ and
$\sigma$ are any two quantum states, then their inner product is
\begin{equation}
  \tr(\rho\sigma) = \sum_{ij} p_\rho(E_i) p_\sigma(E_j)
  \,\tr(\widetilde{E}_i \widetilde{E}_j).
\end{equation}
A convenient fact of frame theory has it that the Gram matrix of the
dual basis is the inverse of the Gram matrix $G$ of the original, so
\begin{equation}
  \tr(\rho\sigma) = \sum_{ij} p_\rho(E_i) p_\sigma(E_j)
  [G^{-1}]_{ij}.
\end{equation}
Note that \emph{if} the inverse Gram matrix $G^{-1}$ could be the
identity, then the Hilbert--Schmidt inner product of density matrices
would reduce to the Euclidean inner product of probability vectors. In
particular, orthogonality of matrices would correspond exactly to
disjoint support of probability distributions. But we know that no MIC
can ever exist for which $G^{-1}$ is the identity matrix~\cite{DeBrota:2018a}.

Trying to make orthogonal quantum states have disjoint probabilistic
representations is asking for a self-dual basis, and no MIC can ever
be self-dual. The closest any MIC can ever come is by being a
SIC~\cite{DeBrota:2018a, DeBrota:2018b}.

Now that we understand the significance of $L$ a bit more, let's
return to the more general context, where the fundamental inequalities
are
\begin{equation}
  L \leq \inprod{p}{s} \leq U
\end{equation}
for probability vectors $p$ and $s$ of Euclidean dimension $N$.  As in
the qplex paper~\cite{Appleby:2017}, we consider mutually maximally
distant (MMD) sets of probability distributions.  An MMD set $\{p_k\}$
satisfies
\begin{equation}
  \inprod{p_k}{p_k} = U,\ \inprod{p_k}{p_{l\neq k}} = L
\end{equation}
for $k,l = 1,\ldots,m$.  Changing to barycentric coordinates
\begin{equation}
  p_k' = p_k - c,
\end{equation}
the shifted vectors satisfy
\begin{equation}
  \inprod{p_k'}{p_k'} = U - \frac{1}{N}
  = r_{\rm out}^2,\ \inprod{p_k'}{p_{l \neq k}'} =
  L - \frac{1}{N} = -r_{\rm mid}^2.
\end{equation}
Construct the sum of all the $\{p_k'\}$:
\begin{equation}
  V = \sum_{k=1}^m p_k'.
\end{equation}
The norm of $V$ is
\begin{equation}
  \inprod{V}{V} = m r_{\rm out}^2 + (m^2 - m)(-r_{\rm mid}^2).
\end{equation}
Because the norm of a vector is always nonnegative, we have that the
size $m$ of the MMD set satisfies the bound
\begin{equation}
  m \leq 1 + \frac{r_{\rm out}^2}{r_{\rm mid}^2}.
\end{equation}

When the value of $m$ attains this upper bound, the norm of $V$ is
zero, and thus $V$ itself must be the zero vector, meaning that
\begin{equation}
  \sum_k p_k = mc.
\end{equation}

The Gram matrix for the MMD set $\{p_k\}$ is
\begin{equation}
  G = (U-L)I_m + LJ_m.
\end{equation}
Using the values familiar from quantum theory, $m \leq d$ and
\begin{equation}
  [G]_{jk} = \frac{1 + \delta_{jk}}{d(d+1)}.
\end{equation}
Because the Gram matrix is invertible, the vectors $\{p_k\}$ are
linearly independent.  Consequently, the only way to get $p_1$ through
$p_m$ to sum up to the vector $mc$ is to add them with equal weights,
as above.  We see that the only way the MMD set $\{p_k\}$ can
constitute a measurement matrix is to set
\begin{equation}
  r(k|i) = \frac{N}{m} p_k(i).
\end{equation}
Given the garbage state $c$, all outcomes of this measurement are
necessarily assigned the same probability $1/m$.

Let $d$ denote the bound we deduced above on the size of an MMD set in
terms of $N$ and $L$. By assuming that this bound is in fact an
integer, we can show that
\begin{equation}
  N = d + \qbar \frac{d(d-1)}{2},
  \label{eq:polygonal}
\end{equation}
where $\qbar$ is a nonnegative integer.  Fixing a value of~$\qbar$
fixes the constants $N$, $L$ and $U$ in terms of the parameter $d$.
Homogeneity of the qplectic cone, the assumption with which we have
gradually made peace, leads us to set $\qbar = 2$, so that $N = d^2$
and $U$ is exactly twice $L$:
\begin{equation}
\frac{1}{d(d+1)} \leq \inprod{p_1}{p_2} \leq \frac{2}{d(d+1)}.
\end{equation}
With this choice, the generalized urgleichung becomes the original
urgleichung stated in earlier work~\cite{Fuchs:2013b}:
\begin{equation}
q(j) = \sum_i\left[(d+1)p(i) - \frac{1}{d}\right]r(j|i).
\end{equation}

I've often wanted to begin a lecture on some abstruse technical topic
by intoning, in my best pompous announcer voice, ``Since Man first
looked up at the stars in wonder, he has asked himself, are all
SIC-POVMs group covariant?''  At this juncture, I would have better
warrant to do so, because our relation between $N$ and $d$ is an
instance of a formula that ``Man'' has known for quite a while.  When
we set $\qbar$ equal to 2, then we get that $N$ is just the square
of~$d$; in other words, $N$ ranges over the square numbers.  If
instead we fix $\qbar$ equal to 1, $N$ will range over the triangular
numbers.  And, for an arbitrary value of~$\qbar$, this formula says
that $N$ is the $d^{\rm th}$ \emph{polygonal number,} where the
polygons in question have $\qbar+2$ sides~\cite{Conway:1996}.
Polygonal numbers go back to Pythagorean number mysticism, and were
studied as long ago as Hypsicl\={e}s, who was active in the second
century BCE.\footnote{When you go back that far, the history of
  mathematics and science becomes semi-legendary. The best one can
  typically do for ``evidence'' is a fragment of a lost book quoted in
  another book that happened to survive, and all of it dating to
  decades or centuries after the events ostensibly being
  chronicled. Did Pythagoras actually prove the theorem we named after
  him, or did he merely observe that it held true in a few special
  cases, like the 3--4--5 right triangle?  Tough to say, but the
  latter would have been easier, and it would seem to appeal to a
  number mystic, for whom it's all about the successive whole
  numbers. Pythagoras himself probably wrote nothing, and nothing in
  his own words survives. It's not clear whether his contemporaries
  viewed him as a mathematician or primarily as a propounder of an
  ethical code. (Even only 150 years after the time he purportedly
  lived, the ancient authorities disagreed about whether Pythagoras
  was a vegetarian, with Aristoxenus saying no and Eudoxus
  yes~\cite{Huffman:2014}.) Suppose that Pythagoras had never lived,
  and a cult had attributed their work to that name in ritual
  self-denial --- like the Bourbaki collective~\cite{Richer:2013}, we
  might say, but more so.  Their bibliographic practices would not be
  exactly the same as ours today.  Where we'd say, ``This idea comes
  from Egyptian mathematics, where it is stated in the Rhind
  papyrus,'' they might have said, ``Pythagoras learned this idea in
  Egypt.''  Later, parables of this kind could have been taken for
  biography: ``In his youth, Pythagoras visited Egypt.''  The result
  of such a process would be would be hard to tell from the surviving
  historical evidence we have today.}

The Pythagorean number mystics could have arrived at the rule we
express by Eq.~(\ref{eq:polygonal}) from such a simple starting point
as arranging pebbles into nice shapes and then counting how many
pebbles the shape contains.  Shapes of larger and larger size are
built up by adding more pebbles.  The arrangement of pebbles added in
each step is called the \emph{gnomon,} and it has the form of $\qbar$
sides of a regular polygon with~$(\qbar + 2)$ sides total.  For
example, by starting with a single pebble, and then stacking a
vertical line of two pebbles next to that, three pebbles next to that,
and so on, we build up triangles, whose pebble populations are sums of
the natural numbers:
\begin{equation}
1 + 2 + \cdots + d = \frac{d(d+1)}{2}.
\end{equation}
If we instead build our shape outwards by adding two sides of a
square, we get the square numbers.  If the gnomon has $\qbar$ sides,
then the increment between successive polygonal numbers grows
by~$\qbar$ with each step.  For a given value of~$\qbar$, the $d^{\rm
  th}$ polygonal number is the sum of the first $d$ terms of the
sequence that starts with~1 and grows by~$\qbar$ at each step:
\begin{equation}
N = \sum_{k=0}^{d-1}(1 + k\qbar) = d + \qbar \sum_{k=0}^{d-1} k.
\end{equation}
Because the second term is just $\qbar$ times the $(d-1)^{\rm th}$
triangular number, we can also arrange our $N$ pebbles as a line
of~$d$, plus $\qbar$ triangles of side length $d-1$ pebbles
each.\footnote{The polygonal numbers, as a class, are not the sort of
  mathematics that physics habitually invokes. However, historians of
  science observe that Leonardo da Vinci wrote, ``The freely falling
  body acquires with each unit of time a unit of motion, and with each
  unit of motion a unit of velocity''~\cite{Drake:1975}. One reading
  of this is that Leonardo believed that the distance a body falls in
  successive time intervals goes as the positive integers. The
  \emph{total} distance fallen after $T$ time units is thus a
  triangular number. The step to Galileo's law of falling bodies is to
  replace successive integers with successive odd numbers, making the
  total distance fallen go as the perfect squares. Amusingly, this is
  the same change that separates quantum mechanics from its most
  closely-studied foil theory~\cite{Hardy:2012, Wootters:2013,
    Wootters:2013b}.}

Thus, the function $N(d)$ for a given choice of $\qbar > 0$ would be
the number of real parameters in a $d \times d$ self-adjoint matrix
wherein the elements have $\qbar - 1$ distinct imaginary units.  As is
well known, ``numbers'' of such forms can only constitute normed
division algebras if $\qbar$ equals 1, 2, 4 or 8. Pythagoreans would
probably like the fact that the $N(d)$ relations for real, complex and
quaternionic quantum mechanics correspond to triangles, squares and
hexagons, the three polygons that, when drawn regularly, tile the
plane without gaps.  The fact that these are the three possible
regular tilings is another result whose original proof is lost to
legend~\cite[pp.\ 209--10]{Sarton:1952}.  Despite much mulling over
``Arnold trinities''~\cite{Arnold:2014}, I have been unable to invent
a deep significance for this coincidence.  Nor have I been able to
find a solid connection with the \emph{other} appearance of normed
division algebras in this theory, the classification of the sporadic
SICs~\cite{Stacey:2017}.

\section{(En)Tangled Banks:\ Charting the Shorelines of the Hilbert Qplexes}
\label{sec:hilbert-qplex}
\begin{quotation}
\noindent I first presented the result at an APS meeting a
  couple years ago. Charlie Bennett was in the audience and asked,
  ``Is that a 7?'' I said, ``Yep, it's really a 7.'' Charlie said,
  ``Well then, it's the first 7 I've ever seen in quantum
  information.'' And what else would you expect from a truly
  fundamental equation?! Indeed it is a 7, and well checked many times
  by myself and independently by my students. In fact, just the other
  day by the latest, Ryan Morris, who first found a 6 instead
  \ldots\ but then ultimately found a 7.
  \begin{flushright}--- Chris Fuchs, in correspondence, 2008~\cite{Fuchs:2014}
  \end{flushright}
\end{quotation}

The last technical discussion in this essay will be a more detailed
treatment of the Hilbert qplexes. For a given dimension $d$, a Hilbert
qplex lives in the space $\bbR^{d^2}$, and more specifically in the
hyperplane through that space comprising the vectors whose entries sum
to unity. Like all qplexes, the Hilbert qplexes are closed and
convex. One way to specify a Hilbert qplex is to say it is the convex
hull of those probability distributions that stand for maximal
certainty. These distributions are those which satisfy two conditions,
one quadratic and the other cubic. Given my background, they look like
conditions on diversity indices, or on expected scores in peculiar
games~\cite{Stacey:2015, Leinster:2012, Stacey:2017b}.  The first is
just the upper bound of the fundamental inequalities, and it
demarcates a sphere:
\begin{equation}
  \sum_j p(j)^2 = \frac{2}{d(d+1)}.
\end{equation}
The second --- we sometimes call it the QBic equation --- carves away
at that sphere:
\begin{equation}
  \sum_{jkl} p(j)p(k)p(l) C_{jkl} = \frac{d+7}{(d+1)^3},
  \label{eq:qbic}
\end{equation}
where the tensor $C_{jkl}$ brings in the SICs:
\begin{equation}
  C_{jkl} = {\rm Re}\, \tr \Pi_j \Pi_k \Pi_l.
\end{equation}
Perhaps the easiest way to see that peculiar 7 arise is to note that
in any dimension $d$, we can always construct a \emph{quasi-SIC}, a
set of $d^2$ Hermitian operators that sum to $dI$ and satisfy
\begin{equation}
  \tr Q_j Q_k = \frac{d\delta_{jk} + 1}{d+1},
\end{equation}
while not necessarily being positive
semidefinite~\cite{Appleby:2017}. Any quasi-SIC establishes a mapping
from a qplex to operator space.

Substituting in one of the basis distributions
\begin{equation}
  e_a(j) = \frac{1}{d(d+1)} + \frac{\delta_{aj}}{d+1},
  \label{eq:basis}
\end{equation}
we find that
\begin{equation}
  \sum_{jkl} e_a(j) e_a(k) e_a(l) {\rm Re}\,\tr Q_jQ_kQ_l
  = \frac{d + 6 + {\rm Re}\, \tr Q_a^3}{(d+1)^3}.
  \end{equation}
The ``first 7 in quantum information'' occurs when $\tr Q_a^3$ is
constant and maximized, which happens when the quasi-SIC is a genuine
SIC.\footnote{Incidentally, evaluating the sum in Eq.~(\ref{eq:basis})
  appears to be one of the recurring common patterns that the ancient
  Egyptians relied upon in their arithmetic~\cite{Reimer:2014}.}

Asking that the contraction of $C_{jkl}$ with the probability vectors
always evaluates to the same thing is, it seems, a way of asking for
any pure state to be an element in some SIC. That is, while we got
here by imposing homogeneity on the \emph{interior} points, the QBic
equation is telling us about \emph{transitivity} on the \emph{extreme}
points. The shoreline of a Hilbert qplex is made by the orbit of a
single volumeless grain of sand.

The properties of the $C_{jkl}$ tensor that we just invoked follow
from the fact that the $\{\Pi_j\}$ are Hermitian, and more
particularly, that they are rank-1 projectors. We can say much more,
actually:\ The \emph{triple products}
\begin{equation}
  T_{jkl} = \tr \Pi_j \Pi_k \Pi_l
\end{equation}
for any SIC necessarily have a rather rich algebraic
personality~\cite{Appleby:2011, Appleby:2015}. Furthermore, for all
the SICs discovered to date, they also reach out to other areas of
mathematics, sometimes group theory and discrete
geometry~\cite{Stacey:2016b}, and sometimes algebraic number
theory~\cite{Bengtsson:2017, Appleby:2017b}, and nobody knows why. I
suspect there is even more to tell than that. All the SICs known so
far enjoy a property called \emph{group covariance} (and again, nobody
knows why). That is, each SIC can be constructed by starting with an
initial vector $\ket{\pi_0}$ and taking its orbit under the action of
a group, and so that group can serve to turn any $\ket{\pi_j}$ into
any other $\ket{\pi_k}$ in the SIC. Moreover, in almost all cases, the
group is the \emph{Weyl--Heisenberg group} for dimension $d$; the
exception is a class of solutions in $d = 8$ which use a close
variant, and which relate, as so many mathematical exceptions seem to
do, with the octonions~\cite{Stacey:2017}. The Weyl--Heisenberg group
is of considerable technical and historical
importance~\cite{Fuchs:2017}, and so its tight connection with SICs
warrants much thinking upon.  About the best we can currently say in
general terms is Huangjun Zhu's result that group covariance in prime
dimension implies Weyl--Heisenberg covariance~\cite{Zhu:2010b}.

Let $V$ be a Weyl--Heisenberg operator, and consider the
\emph{Clifford group}, the normalizer of the Weyl--Heisenberg group
--- that is, the set of unitaries that map the set of Weyl--Heisenberg
operators to itself up to phase factors. Suppose that $\{W(t)\}$ is a
one-parameter family of Clifford unitaries, for convenience chosen
such that no extraneous phase factors arise. The parameter $t$ can be
interpreted as the clock in a discrete time-evolution process.  Then,
the quantity
\begin{equation}
  F(t) = \bra{\pi_0} W(t)^\dag V^\dag W(t) V \ket{\pi_0}
\end{equation}
will always be the inner product between $\ket{\pi_0}$ and some SIC
vector $\ket{\pi_j}$ (possibly multiplied by an overall root of unity):
\begin{equation}
  \braket{\pi_j}{\pi_0} = \frac{e^{i\theta_j}}{\sqrt{d+1}},\ j \neq 0.
\end{equation}
These numbers are closely related to the triple products, because
\begin{equation}
  \tr \Pi_j \Pi_k \Pi_l = \braket{\pi_j}{\pi_k}
  \braket{\pi_k}{\pi_l}
  \braket{\pi_l}{\pi_j}.
\end{equation}
When all three indices are distinct, this will be a complex number of
magnitude $(d+1)^{-3/2}$ and phase $e^{i\theta_{jkl}}$, which we can
interpret as a \emph{geometric phase} factor~\cite{Mukunda:2002}. And
the quantity $F(t)$ is, formally, equal to an \emph{out-of-time-ordered
  correlator} (OTOC) for the Clifford time evolution $W(t)$. OTOCs
have become of considerable interest in quantum chaos and
thermodynamics~\cite{Halpern:2017, Halpern:2018}. Loosely adapting the
language one hears in APS talks about OTOCs, we might say that the
geometric phases indicate how quantum information is redistributed as
we go from one SIC vector to another.

The QBic equation (\ref{eq:qbic}) looks a little like differential
geometry, a little like game theory; and it brings algebraic number
theory into unusual proximity with quantum chaos.

\section{Conclusion}
I have elsewhere devoted an excessive number of words to setting QBism
in its historical context~\cite{Stacey:2016c, DeBrota:2018c,
  Stacey:2018c}. In this essay, I have tried to be more
forward-looking, though this has involved bringing up old things to
make them part of the new.

Borges once wrote an essay where he listed stories that he found
Kafkaesque.  The common denominator of these stories, going back to
Zeno's paradox of Achilles and the tortoise, was that they all felt
like Kafka, but that he would never have noticed a shared thread
between them if he had not read Kafka.  So, Borges argued, Kafka
invented his precursors~\cite{Borges:1951}.

Likewise, if there were ever a Niels Bohr whose philosophy was
compatible with QBism, it would have to be a Bohr that QBism invented,
as Kafka did for Zeno of Elea.

One of the talking heads in \booktitle{The Creation of the Universe}
was John Archibald Wheeler, who spoke of his belief that under it all
there must lie, not a simple equation, but a compellingly simple
idea. He gave a version of the parable that he used in his first-year
classical mechanics course at Princeton~\cite{Fuchs:2014}, where he
asked each student to write down what they thought were the most
important equations of physics.
\begin{quotation}
\noindent He gathered the papers up and placed them all side-by-side
on the stage at the front of the classroom.  Finally, he looked out at
the students and said, ``These pages likely contain all the
fundamental equations we know of physics.  They encapsulate all that's
known of the world.''  Then he looked at the papers and said, ``Now
fly!''  Nothing happened.  He looked out at the audience, then at the
papers, raised his hands high, and commanded, ``Fly!'' Everyone was
silent, thinking this guy had gone off his rocker.  Wheeler said,
``You see, these equations can't fly.  But our universe flies.  We're
still missing the single, simple ingredient that makes it all fly.''
\end{quotation}

Wheeler was once asked, ``Is the Big Bang here?''  Judging by his
response, he found the question rather
charming~\cite{Wheeler:1982}. It is of course wholly orthodox to say
that the Big Bang \emph{was} here, as the theory of the metric
expansion of space has detailed~\cite{Durrer:2008}. Withdraw all the
air from an ideal balloon, and all the galaxies drawn on it come
together, as we have each heard often enough. Wheeler took the
question down a different path, stating, ``Each elementary quantum
phenomenon is an elementary act of `fact creation.'\,'' He wondered,
``Have we had the mechanism of creation before our eyes all this time
without recognizing the truth?'' QBism took this idea and ran with
it:\ The Bang was here long before us, \emph{is} here with us,
\emph{will be} here after us. Each quantum measurement is a personal
sampling of it, a species of the ongoing creatia distinguished by the
fact that one participant is an agent who bears expectations and the
burden of choice.

This essay began with a probability simplex and a bilinear form,
notions that are crisply geometrical and briefly stated. Yet the story
of how the resulting structure of maximal symmetry fits into
$\mathbb{C}^d$ explodes into tottering piles of nested radicals ---
chaos within order~\cite{Scott:2010}. Looking more deeply still, we
find hints of an order within that chaos --- the construction of
superlatively symmetric measurements from number
theory~\cite{Bengtsson:2017, Appleby:2017b}.

I recall that years ago, before my adventure into QBism, I read an
argument in what I think was an interview with Anton Zeilinger. In my
memory, he said that measurement is an essential part of doing
science, and so it should not be surprising that the next great
advance in fundamental physics might require an improved theory of the
measurement concept. I cannot find this interview, however, and it is
possible that I am remembering the attribution incorrectly. In any
case, whoever said it, I think it is a much more healthy attitude than
rejecting the very idea that a theory of measurement can be a theory
of physics. This latter position is, of course, an article of faith to
many. I might well have had more sympathy myself for this creed in my
younger days, had I spent much time thinking about the issue at
all. For in those days, I was less experienced in the reality of
physical practice, and I was temperamentally inclined to align myself
with whoever proclaimed their love of physics the most loudly. My
youthful affinity for this way of making judgments was, I'd say, very
much akin to my unexamined belief that our profession is a
meritocracy, naturally immune to discrimination by race or by
gender~\cite{PrescodWeinstein:2017, Barres:2018, Wade:2018,
  Miller:2019, Aycock:2019, Rudolph:2019, Harmon:2019, Roebke:2019,
  Watson:2019}. I must plead my youth, not as excuse but only as
explanation.

We physicists can be great hypocrites. For instance, it is common to
complain about the imprecision of the word \emph{measurement} in
quantum mechanics, particularly as older texts use
it~\cite{Chevalley:1999, Stacey:2019}. But you will seldom hear a
physicist make a peep about the word \emph{event} going undefined in a
probability book, or for an example even more within our wheelhouse,
in a text on special relativity. In a homework problem for special
relativity, an \emph{event} could be a single electron annihilating
with a positron, or it could be the explosion of a supergiant
star. This is, I'm sure you'll agree, a category with no well-defined
boundaries. It is criminally vague. One wonders if we could somehow
blame Niels Bohr for it.

As a profession, we have found it adequate to gloss an \emph{event} in
special relativity as a phenomenon whose spatial extent and temporal
duration can be neglected for the purposes at
hand~\cite{Mermin:2018}. To be a touch more technical about it,
\emph{events} are phenomena that we can associate with points in the
conceptual contrivance we call Minkowski spacetime. (``A point,'' we
learned from Euclid, ``is that which has no parts''~\cite{Byrne:1847};
yet a train pulling into a station or a clock striking noon definitely
has parts in plenty.) Saying that a \emph{measurement} on a quantum
system is a process that we can associate with a set of positive
semidefinite operators is no less respectable. The only reason we ever
had to think otherwise was the historical accident that we could see
how to deduce the Minkowski metric from Einstein's
postulates~\cite{Mermin:2005} before we had an equally principled
construction of the quantum formalism. \qed

\bigskip

This research was supported by the John Templeton Foundation. The
opinions expressed in this publication are those of the author and do
not necessarily reflect the views of the John Templeton
Foundation. Among many correspondents, collaborators and
co-conspirators, I would like to extend special thanks to Marcus
Appleby and Howard Barnum, who carried on e-mail discussions during
quite distracting times. Furthermore, I am profoundly grateful to John
C.\ Baez, who gave me the opportunity to write about sporadic SICs and
exceptional Lie algebras at the $n$-Category
Caf\'e~\cite{Stacey:2019c}.  This essay is, I rather suspect, the
counterweight to whatever respectability I gained thereby.

\end{document}